\documentstyle[12pt]{article}
\newtheorem{theorem}{Theorem}

\newtheorem{lemma}[theorem]{Lemma}
\newtheorem{corollary}[theorem]{Corollary}
\newtheorem{definition}[theorem]{Definition}
\newtheorem{remark}[theorem]{Remark}

\def\R{{\bf R}} 
 
\def\N{{\bf N}}
  
\def\C{{\bf C}}

\def\be{\begin{equation}}
\def\ee{\end{equation}}

\def\ds{\displaystyle}

\def\re{{\rm Re}}
\def\im{{\rm Im}}
\def\RS{Ray\-leigh-Schr\"o\-din\-ger}
\def\Sc{Schr\"odinger}
\date{}
\begin{document}
\baselineskip=18pt
\title{Distributional Borel Summability of Odd Anharmonic
Oscillators\footnotemark\footnotetext{Partially supported by Universit\`a di Bologna.
Funds for selected research topics.}}

\author{
Emanuela Caliceti
 \\Dipartimento di Matematica, Universit\`{a} di Bologna
\\40127 Bologna, Italy}
\maketitle 
\vskip 12pt\noindent 
\begin{abstract}
 { \noindent
It is proved that the divergent Rayleigh-Schr\"odinger perturbation expansions for the
eigenvalues of any odd anharmonic oscillator are Borel summable in the distributional
sense to the resonances naturally associated with the system.}
\end{abstract}
\vskip 12pt\noindent 
\section{Introduction and statement of the results}
\setcounter{equation}{0}%
\setcounter{theorem}{0}%
Recent work on complex operators with real spectrum (see e.g.
\cite{Bender1,BeDu,BJK,Cannata,DePh1,DePh2,Zno} and references therein) in quantum
mechanics and on the so-called Bessis-Zinn Justin conjecture have generated a renewed
interest on spectral and  perturbation theory of odd anaharmonic oscillators in quantum
mechanics, namely the class of \Sc\ operators  in
$L^2(\R)$ defined (on a domain to be specified later) by the action of the differential
operator
\be
\label{1.1}
H(\beta)= p^2+x^2+\beta x^{2k+1} \equiv H(0)+\beta x^{2k+1}, \quad
k=1,2,\ldots 
\ee
Here $p=-id/dx$, $\beta$, the coupling constant, is a numerical parameter and $k$ is 
fixed. The
spectral and perturbation theory of the operators $H(\beta)$  (the first perturbation 
theory
examples even introduced in quantum mechanics: see e.g.\cite{Bo}) was settled long ago,
from a mathematically rigorous standpoint, for non-real values of the coupling constant
(\cite{CGM}; see also
\cite{A1,A2,GaPa}). The main results can be summarized as follows: 
\begin{enumerate}
\item
 If $\beta\in \C$, $\im \beta >0$ (analogous
results for $\im \beta <0$) the operator family 
$H(\beta)$ defined on the maximal domain $\ds D(p^2)\cap
D(x^{2k+1})$ is closed and has
compact resolvents.
\item  
$\forall j=0,1,\ldots$,  $H(\beta)$ admits one and only one eigenvalue
$E_j(\beta)$ near the eigenvalue $2j+1$ of
$H(0)$ for $|\beta|$ suitably small, $\im\beta >0$.
\item   
The function $E_j(\beta)$ is holomorphic for $\im \beta>0$, and 
 admits  a
(many-valued) analytic continuation across the real axis to the (Riemann surface)
sector 
$$
S_1(\delta) = \{\beta : |\beta|<B(\delta), -(2k-1)\pi/8 + \delta < \arg{\beta} < 
(2k+7)\pi/8 - \delta \} \; , \quad \forall \delta >0 \, .
$$ 
\item The
\RS\ perturbation expansion $\ds \sum_{s=0}^{\infty}a_s\beta^s$
near the unperturbed eigenvalue $2j+1$ exists to all orders; it has the
property  $a_{2l+1} = 0$, $\forall l\in\N$, and 
is Borel (more precisely, Borel-Leroy of
order $q \equiv (2k-1)/2)$ summable to $E_j(\beta)$ for $\pi/8 +\eta < \arg{\beta}<
7\pi/8-\eta$, $\eta >0$ (see
\cite{BeDu} for tests of numerical accuracy). In particular this implies that 
if $\beta$ is purely imaginary and small as in Item 1 above the eigenvalues 
$E_j(\beta)$ are real. 
\end{enumerate}
A major problem left completely open by these results is however the 
meaning of the perturbation series  for $\beta\in \R$. In this case the operator
$H(\beta)$ defined on the maximal domain is not self-adjoint; it admits
infinitely many self-adjoint extensions, each one with pure point spectrum (see
e.g.\cite{Reed-Simon}, Vol.II).
 Now the  real part of any function $E_j(\beta), \beta\in\R$, which  has 
  no relation with the eigenvalues of the self-adjoint extensions (\cite{CGM}), 
admits  the \RS\ perturbation expansion as an asymptotic expansion to all
orders. On the other hand, it is the function $E_j(\beta)$ which has a physical
meaning: any such complex eigenvalue can be indeed interpreted as a (limit) resonance
of the problem, because it represents the limit of the sequence of shape resonances
obtained by a general cut-off procedure of the potential at infinity as the cut-off is
removed (\cite{CM}). The function
$\re{E_j(\beta)}$ is thus the natural candidate to represent the Borel sum of the
original, real perturbation series; as in the Stark effect, the function $E_j(\beta)$
itself is the natural candidate to represent both location (by its real part) and width
(by its imaginary part) of the resonance. However, when
$\beta\in\R$ the coefficients of this power series have constant sign; as is well
known, this prevents  Borel summability because the Borel transform develops a pole on
the positive real axis. 

The notion of distributional Borel summmability (more precisely,  in this case, 
Borel-Leroy of order $q$, as recalled in the statement of Theorem
\ref{T.1.1} below) was introduced in \cite{DBS} exactly to deal with this kind of
situations, and its validity was proved in (\cite{Db,Stark}) for the perturbation
expansions of the double well oscillator and of the Stark effect, respectively. In this
last case the distributional Borel summability puts into one-to-one
correspondence the perturbation series near the Hydrogen bound states with the
real part (location) of the resonances.
Here the analogous result is proved for the odd anharmonic oscillators, namely:
\begin{theorem}
\label{T.1.1}
 Let
$q = (2k-1)/2$, $\beta\in\R$, $j\in\N$, and $f_j(\beta)\equiv \re{E_j(\beta)}$,
$g_j(\beta)\equiv \im E_j(\beta)$. Then  the \RS\ perturbation expansion  is Borel-Leroy
summable of order $q$ in the distributional sense to $f_j(\beta)$ for $|\beta|$ suitably small,
i.e.:
\begin{itemize} 
\item[(i)] Set 
\be
\label{1.2}
B_{j}(t)\equiv \sum_{s=0}^{\infty}\frac{a_s}{\Gamma (qs+1)}t^s
\ee
Then $B_{j}(t)$ is holomorphic in some circle $|t|< \Lambda_j$;
moreover $B_{j}(t)$ 
admits a holomorphic continuation to the intersection of some neighbourhood of $\R_+$ 
with ${\C}^+\equiv \{t\in{\C}\,:\,\im t >0\}$.
\item[(ii)]
The boundary value distributions $\ds B_j(t\pm i0)$ exist $\forall t\in\R_+$ and the 
following representation holds:
\be
\label{1.3}
f_j(\beta)=\frac{1}{q|\beta|}\int_0^{\infty}PP(B_j(t))
e^{-(t/|\beta|)^{1/q}}\left(\frac{t}{|\beta|}\right)^{-1+1/q}\,dt
\ee 
where $\ds PP(B_j(t))=\frac12(B_j(t+i0)+\overline{B_j(t+i0)})$.
\item[(iii)] $f_j(\beta)=f_j(-\beta)$, $g_j(\beta)=-g_j(-\beta)$.
\end{itemize}
\end{theorem}
\begin{remark}
\label{R.1.2}
\begin{enumerate}
\item

{\rm As for the ordinary Borel sum, the representation (\ref{1.3}) is unique among
all real functions admitting the prescribed formal power series expansion
and fulfilling suitable analyticity requirements and remainder estimates (the
Nevanlinna conditions: see below for their definition and verification in the
distributional case).}
\item
{\rm The symmetry property $f_j(\beta)=f_j(-\beta)$ is a consequence of the
property $a_{2l+1} = 0$, $\forall l$, which in turn follows from the odd symmetry of
the  perturbation $\ds x^{2k+1}$.}
\item
{\rm The distributional Borel summability procedure actually determines also the
imaginary part of the functions $E_j(\beta)$, $\beta\in \R$, i.e. also the width of the
resonances.  The discussion of this aspect is postponed after the proof of Theorem
\ref{T.1.1}.}
\end{enumerate}
\end{remark}
The proof of Theorem \ref{T.1.1} requires the verification of the
analogous of the Nevanlinna criterion as stated and proved in Theorem 4 of \cite{DBS}.
 This is accomplished in two steps. In the first one (details in Sect.2) it
is proved that the eigenvalues $E_j(\beta)$, $\im\beta >0$, admit a (many-valued)
analytic continuation to a  sector wider than the one obtained in \cite{CGM}
, namely $-(2k-1)\pi/4 < \arg{\beta}<(2k+3)\pi/4$. To do this 
we apply to this situation the Hunziker-Vock technique (\cite{HV}), developed after
\cite{CGM},    
 to establish eigenvalue stability. The 
second one (Section 3) consists in extending this analyticity to a suitable
Nevanlinna disk, as required by the criterion for distributional Borel summability. We
do this by  adapting to the present situation the techniques introduced in
\cite{Db,Stark} to deal  with the double well oscillators and the Stark effect.

\section{Analytic continuation of the complex eigenvalues}
\setcounter{equation}{0}%
\setcounter{theorem}{0}%

Let $k\in\N$ be fixed and $\beta\in\C-\{0\}$; $H(\beta)$ will denote the operator 
in $L^2(\R)$ defined by: $D(H(\beta)) = D(p^2)\cap D(x^{2k+1})$ and
\begin{equation}\label{2.1}
H(\beta)u = (p^2 + x^2 + \beta x^{2k+1})u \; ,\quad \forall u\in D(H(\beta)).
\end{equation}
In \cite{CGM} it was proved that, for $\im \beta >0$, $H(\beta)$ represents a 
holomorphic family of type A of operators with compact resolvents and, 
for $|\beta|<B$, non-empty (discrete) spectrum. The norm resolvent convergence 
of $H(\beta)$ to the harmonic oscillator
\be \label{2.2}
H(0) = p^2 + x^2 \; , \quad D(H(0)) = D(p^2)\cap D(x^2) 
\ee
as $|\beta|\to{0}$, $\im\beta>0$, yielded the stability of the eigenvalues of $H(0)$
with respect to the family $H(\beta)$ in the following sense: for any fixed $j\in\N$ and
$\forall \delta >0$, there exists $B_j(\delta)\equiv B(\delta)>0$ such that for
$|\beta|<B(\delta)$,
$\im \beta >0$, $H(\beta)$ has exactly one eigenvalue $E_j(\beta)$ such that 
$|E_j(\beta) - (2j+1)|<\delta$, and therefore $E_j(\beta) \to (2j+1)$ as 
$|\beta|\to{0}$, $\im\beta>0$. Moreover such eigenvalues are analytic functions of
$\beta$, for $|\beta|<B(\delta)$, $\im\beta>0$, and they admit a (many-valued)
analytic continuation across the real axis to the sector 
\be\label{2.3}
S_1(\delta) = \left\{\beta\in\C : |\beta|<B(\delta), -(2k-1)\frac{\pi}{8}+\delta<
\arg{\beta}<(2k+7)\frac{\pi}{8}-\delta \right\}.
\ee
Finally, there exist constants $C,\eta >0$ such that the corresponding \RS\
perturbation expansion is Borel summable to $E_j(\beta)$ in the sector $|\beta|<C$,
$\pi/8 + \eta <\arg{\beta} <7\pi/8 - \eta$. 
The main result in this section consists in extending the analyticity of the 
eigenvalues of $H(\beta)$ to the wider sector $-(2k-1)\pi/4 + \delta<
\arg{\beta}<(2k+3)\pi/4 - \delta$, as stated in the following
\begin{theorem}
\label{T.2.1}
The eigenvalues $E_j(\beta)$ of $H(\beta)$, $\im\beta>0$, which exist for
$|\beta|<B$, admit a (many-valued) analytic continuation across the real axis to
any sector 
\be\label{2.3bis}
S(\delta) = \left\{\beta\in\C : |\beta|<B(\delta), -(2k-1)\frac{\pi}{4}+\delta<
\arg{\beta}<(2k+3)\frac{\pi}{4}-\delta \right\}\; , \forall \delta >0.
\ee
\end{theorem}
In order to prove this theorem we need some preliminary results based on the
standard method of dilation analyticity (see e.g.\cite{Reed-Simon}, Vol.IV, \S XIII.10).
More precisely we introduce the operator
\be\label{2.4}
H(\beta,\theta) \equiv e^{-2\theta}p^2 + e^{2\theta}x^2 +\beta
e^{(2k+1)\theta}x^{(2k+1)}
\equiv e^{-2\theta}K(\beta,\theta)
\ee
which, for $\theta\in\R$, is unitarily equivalent to $H(\beta)$, $\im\beta>0$, via the 
dilation operator $U(\theta)$ defined by 
$$
(U(\theta)u)(x) = e^{\frac{\theta}{2}}u(e^\theta x)\; ,\quad \forall u\in L^2(\R).
$$
In \cite{CGM} it was proved that, when defined on $D(p^2)\cap D(x^{2k+1})$,
$H(\beta,\theta)$ represents a holomorphic family of type A of operators with
compact resolvents for
$-(2k-1)\pi/8<\arg{\beta}<(2k+7)\pi/8$, $\im\theta =
(\pi/2 - \arg{\beta})/(2k+3)$. This was obtained by means of a
quadratic estimate for the operator $p^2 + e^{4\theta}x^2 + i|\beta|x^{2k+1}$ (which
corresponds to $K(\beta,\theta)$ for $\arg{\beta} + (2k+3)\im\theta=\pi/2 $),   
valid for $-\pi/2<4\im\theta <\pi/2 $. Now, a first step in the proof of 
Theorem \ref{T.2.1} consists in proving an analogous quadratic estimate for the
operator 
\be\label{2.5}
K(\beta,\theta) = p^2 + e^{4\theta}x^2 + |\beta|e^{i\arg{\beta}+(2k+3)\theta}x^{2k+1}
\ee
under two more general conditions:
\be
\label{2.6}
\left\{\begin{array}{l} 0 < \arg{\beta} + (2k+3)\im\theta < \pi \\
0 < \arg{\beta} + (2k-1)\im\theta < \pi \end{array}\right.
\ee
\begin{remark}
\label{R.2.2}
{\rm The first of the (\ref{2.6}) corresponds to require the positivity of the
imaginary part of the coefficient of $x^{2k+1}$; as for the second one, if we denote
$\alpha =
\arg{\beta}+  (2k+3)\im\theta$ the argument of the coefficient of $x^{2k+1}$, it is
equivalent to require that the coefficient $\gamma\equiv e^{4\theta}$ of $x^2$ is in the
half-plane 
$-\pi + \alpha <\arg{\gamma}< \alpha$.}
\end{remark}
\begin{lemma}
\label{L.2.3}
Let $\alpha \in]0,\pi[$ and $\Omega \subset\C$ be a compact subset of the half-plane
$-\pi + \alpha <\arg{\gamma}< \alpha$. Then there exist $a,b>0$ such that
\be
\label{2.7}
\|p^2u\|^2 +|\gamma|^2\|x^2u\|^2 + |\beta|^2\|x^{2k+1}u\|^2 \leq a\|(p^2 + \gamma x^2 +
|\beta|e^{i\alpha} x^{2k+1})u\|^2 + b\|u\|^2\; ,
\ee
$\forall u\in D(p^2)\cap D(x^{2k+1})$, $\gamma\in\Omega$, $0<|\beta|\leq 1$, $a$ and
$b$ independent of $\gamma$ in $\Omega$ and $\alpha$ in a closed interval comtained in 
$]0,\pi[$. 
\end{lemma}
{{\it Proof.}} We shall prove the following estimate, equivalent to (\ref{2.7}):
\be
\label{2.8}
\|p^2u\|^2 +|\sigma|^2\|x^2u\|^2 + |\beta|^2\|x^{2k+1}u\|^2 \leq a\|(e^{-i\alpha}p^2 +
\sigma x^2 + |\beta| x^{2k+1})u\|^2 + b\|u\|^2\; ,
\ee
$\forall u\in D(p^2)\cap D(x^{2k+1})$, with $\sigma =\gamma e^{-i\alpha}$ varying in a
compact subset of the half-plane $-\pi <\arg{\sigma}< 0$. As quadratic forms on 
$D(p^2)\cap D(x^{2k+1})\otimes D(p^2)\cap D(x^{2k+1})$ we have:
\vskip 0.2cm\noindent
$\ds (e^{i\alpha}p^2 + \overline{\sigma} x^2 + |\beta| x^{2k+1})(e^{-i\alpha}p^2 +
\sigma x^2 + |\beta| x^{2k+1})$
\vskip 0.1cm\noindent
$\ds= (e^{i\alpha}p^2 + |\beta|x^{2k+1})(e^{-i\alpha}p^2+ |\beta|x^{2k+1}) + |\sigma|^2
x^4
+ \re{\sigma} (e^{i\alpha}p^2+|\beta|x^{2k+1})x^2$
\vskip 0.1cm\noindent
$\ds + i\im{\sigma}(e^{i\alpha}p^2+|\beta|x^{2k+1})x^2 
 + \re{\sigma}x^2(e^{-i\alpha}p^2+|\beta|x^{2k+1})
-i\im{\sigma}x^2(e^{-i\alpha}p^2+|\beta|x^{2k+1})$
\vskip 0.1cm\noindent 
$\ds = \left|\frac{\re{\sigma}}{\sigma}\right|(e^{i\alpha}p^2+|\beta|x^{2k+1}
\pm |\sigma|x^2)(e^{-i\alpha}p^2+|\beta|x^{2k+1} \pm |\sigma| x^2)$
\vskip 0.1cm\noindent
$\ds + \left(1 -\left|\frac{\re{\sigma}}{\sigma}\right|\right)
\left[(e^{i\alpha}p^2+|\beta|x^{2k+1})(e^{-i\alpha}p^2+
|\beta|x^{2k+1}) + |\sigma|^2 x^4\right]$
\vskip 0.1cm\noindent
$\ds + i\im{\sigma}(e^{i\alpha}p^2x^2 - e^{-i\alpha}x^2p^2)$
\vskip 0.1cm\noindent
$\ds \geq \left(1 -\left|\frac{\re{\sigma}}{\sigma}\right|\right)
\left[(e^{i\alpha}p^2+|\beta|x^{2k+1})(e^{-i\alpha}p^2+
|\beta|x^{2k+1}) + |\sigma|^2 x^4\right] $
\vskip 0.1cm\noindent
$\ds + i\im{\sigma} \cos{\alpha}[p^2,x^2] - \im{\sigma}\sin{\alpha}(p^2x^2 + x^2p^2)$
\vskip 0.1cm\noindent
$\ds = \left(1 -\left|\frac{\re{\sigma}}{\sigma}\right|\right)
\left[(e^{i\alpha}p^2+|\beta|x^{2k+1})(e^{-i\alpha}p^2+
|\beta|x^{2k+1}) + |\sigma|^2 x^4\right] $
\vskip 0.1cm\noindent
$\ds + 2\im{\sigma} \cos{\alpha}(px + xp) - \im{\sigma}\sin{\alpha}\left([p,[p,x^2]] +
2px^2p\right)$
\vskip 0.1cm\noindent
$\ds = \left(1 -\left|\frac{\re{\sigma}}{\sigma}\right|\right)
\left[(e^{i\alpha}p^2+|\beta|x^{2k+1})(e^{-i\alpha}p^2+
|\beta|x^{2k+1}) + |\sigma|^2 x^4\right] $
\vskip 0.1cm\noindent
$\ds -2\im{\sigma} |\cos{\alpha}|(\mp px \mp xp) - \im{\sigma}\sin{\alpha}(-2 + 2px^2p)$
\vskip 0.1cm\noindent
(since $\sin{\alpha}>0$ and $\im{\sigma}<0$)
\vskip 0.1cm\noindent
$\ds \geq \left(1 -\left|\frac{\re{\sigma}}{\sigma}\right|\right)
\left[(e^{i\alpha}p^2+|\beta|x^{2k+1})(e^{-i\alpha}p^2+
|\beta|x^{2k+1}) + |\sigma|^2 x^4\right] $
\vskip 0.1cm\noindent
$\ds -2\im{\sigma} |\cos{\alpha}|\left[(p \mp x)^2 -p^2 -x^2\right] 
+2\im{\sigma}\sin{\alpha}$
\vskip 0.1cm\noindent
$\ds \geq \left(1 -\left|\frac{\re{\sigma}}{\sigma}\right|\right)
\left[(e^{i\alpha}p^2+|\beta|x^{2k+1})(e^{-i\alpha}p^2+
|\beta|x^{2k+1}) + |\sigma|^2 x^4\right] $
\vskip 0.1cm\noindent
$\ds + 2\im{\sigma} |\cos{\alpha}|(p^2 + x^2) + 2\im{\sigma}\sin{\alpha}$.
\vskip 0.2cm\noindent
In \cite{CGM} it was proved that there exist $a_1,b_1 > 0$, in general depending on 
$|\beta|$, such that 
$$ 
(e^{i\alpha}p^2+|\beta|x^{2k+1})(e^{-i\alpha}p^2 + |\beta|x^{2k+1}) \geq a_1(p^4 + 
|\beta|^2x^{4k+2}) - b_1 \; ;
$$ 
thus,
\vskip 0.2cm\noindent
$\ds (e^{i\alpha}p^2 + \overline{\sigma} x^2 + |\beta| x^{2k+1})(e^{-i\alpha}p^2 +
\sigma x^2 + |\beta| x^{2k+1})$
\vskip 0.1cm\noindent
$\ds \geq A(p^4 + |\beta|^2x^{4k+2}) + B|\sigma|^2x^4 + 2\im{\sigma} |\cos{\alpha}|(p^2
+ x^2) + 2\im{\sigma}\sin{\alpha} -b_1$
\vskip 0.1cm\noindent
$\ds \geq [Aa'p^4 + 2\im{\sigma} |\cos{\alpha}|p^2 + 2\im{\sigma}\sin{\alpha} - b + 
b'/2]$
\vskip 0.1cm\noindent
$\ds + [Aa'|\beta|^2x^{4k+2} + 2\im{\sigma} |\cos{\alpha}|x^2 + b'/2]$
\vskip 0.1cm\noindent
$\ds + A(1 - a')p^4 + A(1 - a')|\beta|^2x^{4k+2} + B|\sigma|^2x^4 - b' \; .$
\vskip 0.2cm\noindent
Now it suffices to choose $0<a'<1$ and $b'>0$ such that the  two terms in square  
brackets are positive.
\begin{lemma}
\label{L.2.4}
Let $\beta$ and $\theta$ be fixed, satisfying conditions (\ref{2.6}) and let 
$\alpha = \arg{\beta} + (2k+3)\im{\theta}$, $\alpha \in ]0,\pi[$. Then there exists 
$\xi>0$ such that 
\be
\label{2.11}
\xi\re{\left[e^{-i(\alpha-\frac {\pi}{2})}\left\langle
u,K(\beta,\theta)u\right\rangle\right]} \geq \langle u,p^2u\rangle \; ,\quad \forall
u\in C_0^\infty (\R)\; .
\ee
\end{lemma}
{{\it Proof.}} We have
\vskip 0.2cm\noindent 
$\ds \re{\left[e^{-i(\alpha-\frac {\pi}{2})}\left\langle u,(p^2 + e^{4\theta}x^2 +
|\beta|e^{(2k+3)\re{\theta} + i\alpha}x^{2k+1})u\right\rangle\right]} $
\vskip 0.1cm\noindent
$\ds = \cos{(\alpha-\pi/2)} \langle u,p^2u\rangle +
e^{4\re{\theta}}\cos{(\pi/2 - \alpha + 4\im{\theta})}\langle u,x^2u\rangle $
\vskip 0.1cm\noindent
$\ds + |\beta|e^{(2k+3)\re{\theta}}\cos{(\pi/2)}\langle u,x^{2k+1}u\rangle $
\vskip 0.1cm\noindent
$\ds = \sin{\alpha}\langle u,p^2u\rangle + e^{4\re{\theta}}\sin{(\alpha-
4\im{\theta})}\langle u,x^2u\rangle $
\vskip 0.1cm\noindent
$\ds \geq\sin{\alpha}\langle u,p^2u\rangle $ ,
\vskip 0.2cm\noindent
since $\sin{(\arg{\beta} + (2k-1)\im{\theta})} > 0$ by the second of (\ref{2.6}).
Moreover, since $0<\alpha <\pi$, the lemma is proved with $\xi = (\sin{\alpha})^{-1}$.
\begin{theorem}
\label{T.2.5}
Let $s=\arg{\beta}$ and $t=\im{\theta}$. Then $H(\beta,\theta)$ is a holomorphic family 
of type A of closed operators on $D(H(\beta,\theta) = D(p^2)\cap D(x^{2k+1})$ with
compact resolvents for $\beta$ and $\theta$ such that $s$ and $t$ vary in the
parallelogram $P$ of the $(s,t)$-plane defined by
\be\label{2.12}
P = \{(s,t)\in\R^2 : 0 < (2k-1)t + s < \pi, 0 < (2k+3)t + s < \pi \} \; .
\ee
\end{theorem}
{{\it Proof.}} Lemma \ref{L.2.3} guarantees that $H(\beta,\theta)$ is closed on a domain
independent of $\beta$ and $\theta$ for $\arg{\beta} = s$ and $\im{\theta} =t$
satisfying conditions (\ref{2.6}):
$$
\left\{\begin{array}{l} 0 < (2k+3)t + s < \pi \\
0 < (2k-1)t + s  < \pi \end{array}\right.
$$
which define the parallelogram $P$ with vertices in the points of coordinates
$(-(2k-1)\pi/4,\pi/4)$, $(0,0)$, $((2k+3)\pi/4 ,-\pi/4)$,
$(\pi,0)$. From Lemma \ref{L.2.4} it follows that, for $\beta$ and $\theta$ in this
region, $K(\beta,\theta)$ has numerical range in the half-plane $-\pi + \alpha
\leq \arg{z}\leq \alpha$, with $\alpha = \arg{\beta} + (2k+3)\im{\theta}$; thus  
$H(\beta,\theta)$ has numerical range contained in the half-plane 
$$
\Pi = \{z\in\C : -\pi + \arg{\beta} + (2k+1)\im{\theta} \leq \arg{z} \leq \arg{\beta} + 
(2k+1)\im{\theta}\}.
$$
By standard arguments on the holomorphic families of type A (see \cite{Kato} or \cite
{Reed-Simon} Vol.IV), taking into account the above mentioned results obtained in
\cite{CGM} for 
$-(2k-1)\pi/8<\arg{\beta}<(2k+7)\pi/8 $, we now obtain the analyticity
of $H(\beta,\theta)$ in the region defined by $P$, which allows $\beta$ to be extended
to the sector $-(2k-1)\pi/4 <\arg{\beta}<(2k+3)\pi/4 $, as well as the
compactness of the resolvents. Finally, the (discrete) spectrum of $H(\beta,\theta)$ 
is contained in $\Pi$ and $\forall z\notin\Pi$, $\|(z - H(\beta,\theta))^{-1}\|\leq 
({\rm {\rm dist}}(z,\Pi))^{-1}$.
\begin{remark}
\label{R.2.6}
{\rm Let us notice that, if we start from the operator $H(\beta)$ with $\im{\beta}<0$,
analogous results can be obtained for the operator family $H(\beta,\theta)$ for 
$\beta$ and $\theta$ such that $s = \arg{\beta}$, $t = \im{\theta}$ vary in the
parallelogram
$$
P^1 = \{(s,t)\in\R^2 : -\pi < (2k-1)t + s < 0, -\pi < (2k+3)t + s < 0 \} .
$$
Furthermore the adjoint operator $H(\beta,\theta))^*$ of $H(\beta,\theta)$ is
$H(\overline{\beta},\overline{\theta})$.}
\end{remark}
In order to complete the proof of Theorem \ref{T.2.1} we need to extend to the wider
sector $S(\delta)$ given by (\ref{2.3bis}) the result obtained in \cite{CGM} for $\beta
\in S_1(\delta)$ (see (\ref{2.3})), on the existence of eigenvalues of
$H(\beta,\theta)$ and on their convergence to the corresponding eigenvalues of the
harmonic oscillator as $|\beta|\to{0}$. To this end, since we cannot make use of the
norm resolvent convergence which holds only for $\beta\in S_1(\delta)$,
$|\beta|\to{0}$, we will apply the more general criterion for the stability of the
eigenvalues introduced in \cite{HV} and based on the strong convergence of the
resolvents. More precisely, let us consider the operator
$$
H(0,\theta) \equiv e^{-2\theta}p^2 + e^{2\theta}x^2 \; , \quad D(H(0,\theta)) =
D(p^2)\cap D(x^2)
$$
corresponding to the dilated harmonic oscillator. We will prove that the eigenvalues of 
$H(0,\theta)$, independent of $\theta$ for $-\pi/4 <\im{\theta}<\pi/4 $,
and represented by the sequence of the odd numbers $\{(2j+1) : j\in\N\}$, are stable in
the sense of Kato with respect to the family $\{H(\beta,\theta) : |\beta|>0\}$, $\beta$
and $\theta$ in the region defined by $P$. For simplicity we will work with the
operators $K(\beta,\theta) = e^{2\theta}H(\beta,\theta)$ and $K(0,\theta) =
e^{2\theta}H(0,\theta)$; moreover, from now on we will assume $\theta$ purely
imaginary, that is of the form $i\theta$, $-\frac{\pi}{4}<\theta<\frac{\pi}{4}$, and
(with slight abuse of notation) we will still denote $H(\beta,\theta)$ and
$K(\beta,\theta)$ the operators $H(\beta,i\theta)$ and $K(\beta,i\theta)$ respectively.
Notice that with this convention we should read $\theta$ in place of $\im{\theta}$
wherever the notation $\im{\theta}$ has been employed, in particular in the conditions
(\ref{2.6}). Finally, let
$\sigma(K(\beta,\theta))$ denote the spectrum of
$K(\beta,\theta)$. Then, in order to obtain the above mentioned stability result, we
will prove the following
\begin{theorem}
\label{T.2.7}
Let $\beta$ and $\theta$ satisfy conditions (\ref{2.6}). We have:
\begin{itemize}
\item[(i)]
if $\lambda\notin\sigma(K(0,\theta))$, then $\lambda\in\Delta$, where
$$
\Delta = \{z\in\C : z\notin\sigma(K(\beta,\theta))\; {\rm and} \; (z -
K(\beta,\theta))^{-1}\; {\rm is\; uniformly\; bounded\; as} \; |\beta|\to{0}\}
$$
\item[(ii)]
if $\lambda\in\sigma(K(0,\theta)) = \{(2j+1)e^{2i\theta} : j\in\N\}$, then $\lambda$
is stable with respect to the family $K(\beta,\theta)$, i.e.: if $r>0$ is sufficiently
small, so that the only eigenvalue of $K(0,\theta)$ enclosed in $\Gamma_r = \{z\in\C :
|z-\lambda| = r \}$ is $\lambda$, then there is $B>0$ such that for $|\beta|<B$, 
$dim P(\beta,\theta) = dim P(0,\theta)$, where 
$$
P(\beta,\theta) = (2\pi i)^{-1}\oint_{\Gamma_r}(z - K(\beta,\theta))^{-1}dz
$$
is the spectral projection of $K(\beta,\theta)$ corresponding to the part of the
spectrum enclosed in $\Gamma_r \subset\C-\sigma(K(\beta,\theta))$. Similarly for 
$P(0,\theta)$.
\end{itemize}
\end{theorem}
{{\it Proof.}}
It is a straightforward application of Theorem 5.4 of \cite{HV} once we have proved the
following
\begin{theorem}
\label{T.2.8}
Let $\arg{\beta}$ and $\theta$ be fixed, satisfying conditions (\ref{2.6}), and let
$K(\rho) = K(\beta,\theta)$ with $\rho = |\beta|$. Then
\begin{itemize}
\item[(a)]
$$
\lim_{\rho \rightarrow 0^+} K(\rho)u = K(0)u  , \quad \lim_{\rho \rightarrow 0^+}
K(\rho)^*u = K(0)^*u  ,\quad \forall u\in C_0^\infty(\R).
$$
\item[(b)]
$\Delta \neq \emptyset$.
\item[(c)]
Let $\chi\in C_0^\infty(\R)$ be such that $\chi (x)=1$ for $|x|\leq 1$, $0\leq \chi(x)
\leq 1$, $\forall x\in\R$, $\chi(x)=0$ for $|x|\geq 2$. For $n\in\N$ let $\chi_n(x) = 
\chi(x/n)$ and $M_n(x) = 1 - \chi_n(x)$. We have:
\begin{itemize}
\item[(1)]
if $\rho_m\to 0^+$ and  $\;u_m\in D(K(\rho_m))$ are two sequences such that
$$
\|u_m\|\to 1 , \; u_m\stackrel{w}{\rightarrow} 0 , \quad and \quad \|K(\rho_m)u_m\|\leq
({\rm const.}) ,
\;
\forall m ,
$$
then there exists $a>0$ such that
$$
\limsup_{m\rightarrow\infty} \|M_nu_m\| \geq a > 0 ,\quad \forall n \, ;
$$
\item[(2)]
for some $z\in\Delta$
$$
\lim_{n\rightarrow\infty} \|[M_n,K(\rho)](z - K(\rho))^{-1}\| = 0 \; ,
$$
uniformly as $\rho\to 0^+$;
\item[(3)]
$\forall \lambda\in\C$, there exists $\delta>0$ such that
$$
d_n(\lambda,\rho) \equiv \inf{\{\|(\lambda - K(\rho))M_nu\| : u\in D(K(\rho)),
\|M_nu\|=1\}} > \delta \; ,
$$
$\forall n>n_0$ and $\rho\to 0^+$.
\end{itemize}
\end{itemize}
\end{theorem}
{{\it Proof.}}
\begin{itemize}
\item[{\it (a)}]
It follows immediately from the convergence of the potential $V(\rho) = e^{4i\theta}x^2
+ \rho e^{i(\arg{\beta}+(2k+3)\theta)}x^{2k+1}$ to $V(0) = e^{4i\theta}x^2$ as 
$\rho\to 0^+$, uniformly on the compact subsets of $\R$.
\item[{\it (b)}]
As already observed in the proof of Theorem \ref{T.2.5} $K(\rho)$ has numerical range
contained in the half-plane 
$$
\Pi_\alpha = \{z\in\C : -\pi + \alpha \leq arg{z} \leq \alpha\},\quad \alpha =
\arg{\beta} + (2k+3)\theta
$$
indipendent of $\rho$, and $\forall
z\notin\Pi_\alpha$, $\|(z - K(\rho))^{-1}\| \leq ({\rm {\rm dist}}(z,\Pi_\alpha))^{-1}$.
\item[{\it (c)}]
Statement (1) follows from a standard argument based on an estimate which comes from
Lemma \ref{L.2.4}: there exists $c>0$ such that 
\be\label{2.13}
\|(1 + p^2)^\frac{1}{2}u\| \leq c(\|K(\rho)u\| + \|u\|) \; ,\quad \forall u\in
D(K(\rho)) \, .
\ee
For the details see \cite{HV}. As for (2), following again \cite{HV}, we have:
$$
[M_n,K(\rho)] = [\chi_n,p^2] = 2in^{-1}\Phi_np - n^{-2}\Psi_n \; ,
$$
where the functions $\Phi_n$ and $\Psi_n$, obtained by differentiating $\chi$ once and
twice respectively, are uniformly bounded in $n$ and $\rho$. Thus, the result follows
applying again (\ref{2.13}). Finally, given $\lambda\in\C$ we have
$$
d_n(\lambda,\rho) = \inf{\{\|(\lambda' - e^{i(\frac{\pi}{2}-\alpha)}K(\rho))M_nu\| :
u\in D(K(\rho)), \|M_nu\|=1\}}
$$
with $\lambda' = e^{i(\frac{\pi}{2}-\alpha)}\lambda$, $\alpha = \arg{\beta} +
(2k+3)\theta$. Therefore $d_n(\lambda,\rho) \geq {\rm dist}(\lambda',G_n(\rho))$, where 
$$
G_n(\rho) = \left\{\left\langle M_nu,e^{i(\frac{\pi}{2}-\alpha)}K(\rho))M_nu
\right\rangle :  u\in D(K(\rho)), \|M_nu\|=1\right\} \; ,
$$  
whence
$$
d_n(\lambda,\rho) \geq \inf{\left\{\re{\left\langle
M_nu,e^{i(\frac{\pi}{2}-\alpha)}K(\rho))M_nu\right\rangle} - |\lambda'| : u\in
D(K(\rho)), \|M_nu\|=1\right\}} \; .
$$
Now the assertion follows from the proof of Lemma \ref{L.2.4}, which yields
\vskip 0.2cm\noindent
$\ds \re{\left\langle M_nu,e^{i(\frac{\pi}{2}-\alpha)}K(\rho))M_nu\right\rangle}$
\vskip 0.1cm\noindent 
$\ds \geq \sin{(\arg{\beta} + (2k-1)\theta)}\langle M_nu,x^2M_nu\rangle \geq
n^2\sin{(\arg{\beta} + (2k-1)\theta)}$
\vskip 0.2cm\noindent
and therefore
$$
\lim_{\stackrel{n\to\infty}{\rho\to {0^+}}} d_n(\lambda,\rho) = +\infty \, . 
$$
\end{itemize}
\begin{remark}
\label{R.2.9}
{\rm It is immediate to check that all the results so far obtained, in particular the
analyticity of the family $H(\beta,\theta)$ and the stability of the eigenvalues of the
harmonic oscillator with respect to $H(\beta,\theta)$ as $\rho=|\beta|\to {0^+}$, hold
uniformly in $\beta$ and $\theta$ such that $(\arg{\beta},\theta)$ varies in any
compact subset of $P$.}
\end{remark}
{{\it Proof of Theorem \ref{T.2.1}}}
It follows from Theorems \ref{T.2.5} and \ref{T.2.7} and from Remark \ref{R.2.9}. In
particular if $(\arg{\beta},\theta)\in P$, by the well-known Symanzik scaling
properties (see
\cite{Simon}) the eigenvalues $E_j(\beta)$ of $H(\beta,\theta)$ do not depend on
$\theta$ and represent the analytic continuation to the sector $S(\delta)$ of the
eigenvalues of $H(\beta)$, $\im\beta >0$; in fact, as already observed, the condition 
$(\arg{\beta},\theta)\in P$, allows us to extend $\arg{\beta}$ to the interval 
$]-(2k-1)\pi/4,(2k+3)\pi/4[$.
\begin{remark}
\label{R.2.10} 
{\rm Let $E_j(\beta)$ denote the generic eigenvalue of $H(\beta)$ for $\im\beta >0$,
which can be analytically continued to the sector $S(\delta)$, and $E_j^1(\beta)$ the
generic eigenvalue of $H(\beta)$ for $\im\beta <0$, which can be analytically continued
to the sector
$$
\overline{S}(\delta) = \left\{\beta\in\C : 0<|\beta|<B(\delta),
-(2k+3)\frac{\pi}{4}+\delta<
\arg{\beta}<(2k-1)\frac{\pi}{4}-\delta \right\} \; .
$$
Then, from Remark \ref{R.2.6} we have $E_j^1(\beta) =
\overline{E_j(\overline{\beta})}$.} 
\end{remark}
\par
\section{Analyticity of the eigenvalues in a Nevanlinna disk and distributional Borel
summability}
\setcounter{equation}{0}%
\setcounter{theorem}{0}%
We begin this section by stating and proving the basic analyticity result needed to
establish the distributional Borel summability.
\begin{theorem}
\label{T.3.1}
Set $q = (2k-1)/2$. For each eigenvalue $E_j(\beta)$, $j\in\N$, of the odd
anharmonic oscillator $H(\beta)$ there exists $R>0$ such that $E_j(\beta)$ is analytic
in the Nevanlinna disk $C_R = \{\beta\in\C : \re{\beta^{-1/q}} \geq R^{-1}\}$ of the 
$\beta^{1/q}$-plane.
\end{theorem}
\begin{remark}
\label{R.3.2}
\begin{itemize}
\item[{\rm (I)}]
{\rm the sector $S(\delta)$ can be re-written in terms of the parameter $q$}:
$$
S(\delta) = \left\{\beta\in\C : |\beta|<B(\delta),-\frac{\pi}{2}+\frac{\delta}{q}<
\arg{\beta^{1/q}}<\frac{\pi}{2}+\frac{\pi}{q}-\frac{\delta}{q}\right\} \; .
$$
\item[{\rm (II)}]
{\rm The function $E_j(\beta)$, analytic in any sector $S(\delta)$ and for which we
want to prove analyticity in a disk $C_R$, represents an eigenvalue of the operator 
$H(\beta,\theta)$ if the pair $(\beta,\theta)$ satisfies the condition
$(\arg{\beta},\theta)\in P$. In particular for $\ds -{\pi}(2k-1)/4 < \arg{\beta} <
0$ we can choose the path inside $P$ given by the straight line of equation 
$$ 
\theta = -\frac{1}{2k+1}\arg{\beta} + \frac{\pi}{2(2k+1)} \, ;
$$ 
then, if we set 
$$ 
\arg{\beta} = 
-\frac{\pi}{4}(2k-1) + \frac{\epsilon}{2}(2k-1) = -\frac{\pi}{2}q + \epsilon q,\;
{\rm i.e.}\;
  \arg{\beta^{1/q}} = -\frac{\pi}{2} + \epsilon, \epsilon\to 0^+
$$
 we obtain 
$ \theta = \pi/4 - (2k-1)\epsilon/[2(2k+1)] = \pi/4 - \epsilon q/(2k+1) $, and the
operator
$H(\beta,\theta)$ takes the form}
$$
A(\rho) = e^{-i(\frac{\pi}{2}-\frac{2k-1}{2k+1}\epsilon)}p^2 +
 e^{i(\frac{\pi}{2}-\frac{2k-1}{2k+1}\epsilon)}x^2 + i\rho x^{2k+1} \; , \quad {\rm
with} \; \rho=|\beta| \, .
$$
\item[{\rm (III)}]
{\rm For $\ds \beta = \rho e^{i\arg{\beta}}$ and $\ds \arg{\beta} =
(-\frac{\pi}{2}+\epsilon)q$, the boundary of $C_R$ has equation
\be\label{3.1}
\sin{\epsilon} = \frac{\rho^{1/q}}{R} \; .
\ee
Since the disk $C_R$ can be regarded as the union of the boundaries of disks of smaller
radius, the proof of Theorem \ref{T.3.1} reduces to a stability argument with respect to
the family $A(\rho)$, as $\rho\to 0^+$, under condition (\ref{3.1}), for the eigenvalues
of a suitable limiting operator, which we proceed to define.}
\end{itemize}
\end{remark}
The argument is similar to the one already developed in \cite{Stark} and \cite{Db} to
obtain analyticity of the eigenvalues for the operators associated with the Stark
effect and the double well oscillators respectively. More precisely, let $D$ denote the
dense subset of $L^2(\R)$ of the functions which are translation analytic in a suitable
strip $|\im{x}|<\eta_0$, for some $0<\eta_0 <1$ (recall that $u\in L^2(\R)$ is
translation analytic for $|\im x|<r$ if $(T_au)(x)=u(x+a)$ admits an $L^2-$valued
analytic continuation to $|\im a|<r$);
$D$ represents a core for
$A(\rho)$.
\begin{definition}
\label{D.3.3}
Let $\eta >0$ be fixed and small. For fixed $a_k >0$, set $\ds x_0 =
-\frac{a_k}{\rho^{1/(2k-1)}}$ and let ${\cal U}$ denote the unitary operator in
$L^2(\R)$  defined by 
$$
({\cal U}\psi)(x) = (\xi_{\rho}'(x))^{\frac{1}{2}}\psi(\xi_{\rho}(x)) \; , \quad \forall
\psi\in D
\, ,
$$
where, for any given $\rho >0$, $\xi_{\rho}\in C^{\infty}(\R)$ satisfies the conditions:
\be
\label{3.2}
\begin{array}{l} \xi_{\rho}(x) = x - i\eta\arctan{\left[{x}/{(1+x^2)^{1/4}}\right]} \; ,
\quad  -x_0\leq x < +\infty \\ \xi_{\rho}(x) = x \; , \qquad \qquad \qquad \qquad
\qquad \qquad x \leq x_0-\eta\end{array}
\ee
and $\im{\xi_{\rho}}(x)$ is monotone in the remaining region. 
\par\noindent
Then the closed operator $H_\rho \equiv {\cal U}A(\rho){\cal U}^{-1}$, unitarily
equivalent to $A(\rho)$ and with the same (discrete) spectrum, has $D_1\equiv{\cal
U}(D)$ as a core, and its action on $D_1$ is given by
\be\label{3.3}
H_\rho u = e^{-i(\frac{\pi}{2}-\frac{2k-1}{2k+1}\epsilon)}\left\{pf_{\rho}^2 p + 4^{-1}
(f_{\rho}^2)''\right\}u + e^{i(\frac{\pi}{2}-\frac{2k-1}{2k+1}\epsilon)}\xi_{\rho}^2 u +
i\rho\xi_{\rho}^{2k+1}u\; , \quad \forall u\in D_1\, ,
\ee
where $f_{\rho}(x) = (\xi_{\rho}'(x))^{-1}$, $\forall x\in\R$.
\end{definition}
\begin{remark}
\label{R.3.4}
{\rm In a similar way we can define the dilated harmonic oscillator, having $D_1$ as a
core:
$$
H_0 u = -i\left\{pf_0^2 p + 4^{-1}
(f_0^2)''\right\}u + i\xi_0^2 u  \; , \quad \forall u\in D_1\, ,
$$
where $f_0(x) = (\xi_0'(x))^{-1}$ and $\xi_0'(x) = x -
i\eta\arctan{\left[{x}/{(1+x^2)^{1/4}}\right]}$, $\forall
\in\R$. In Corollary \ref{3.7} we will prove that $H_0$ is the limit in the strong
resolvent sense of $H_\rho$ as $\rho\to 0^+$. Therefore, as anticipated after Remark
\ref{R.3.2}, the proof of Theorem \ref{T.3.1} consists in obtaining a stability result
for the eigenvalues $E_j = (2j+1)$, $j\in\N$, of $H_0$, which coincide with those of
the harmonic oscillator, with respect to the family $H_\rho$ as $\rho\to 0^+$.}
\end{remark}
Proceeding in analogy with \cite{Stark} and \cite{Db}, this result will be obtained by
proving some preliminary lemmas aimed to verify the hypotheses of Theorem A.1 of
\cite{Db}. This theorem represents a simpler tool for applications, in the context of
the more general stability theory developed by Hunziker and Vock in \cite{HV}. In
particular in the subsequent Lemmas \ref{L.3.5}, \ref{L.3.6}, \ref{L.3.9}, \ref{L.3.10}
and Corollaries
 \ref{C.3.7}, \ref{C.3.8}, we follow the corresponding steps used in \cite{Stark} and
\cite{Db} to obtain similar results, each one adapted to the specific characteristics
of the present problem; we will describe here the relevant details.
\begin{lemma}
\label{L.3.5}
Let $V_\rho (x) = e^{i(\frac{\pi}{2}-\frac{2k-1}{2k+1}\epsilon)}\xi_{\rho}^2(x)  +
i\rho\xi_{\rho}^{2k+1}(x)$. Then for a suitable choice of the constant $a_k >0$ in
Definition \ref{D.3.3} there exist constants $c_1>0$ and $c_2\in\R$ such that
\be\label{3.4}
\re {V_\rho (x)} \geq \frac{c_1}{R} + c_2 \; , \quad \forall x\notin (-n,n)
\ee
$\forall n\geq n_0$, $0<\rho<\rho_0$.
\end{lemma}
{{\it Proof.}}
Set $ \eta(x) = \im{\xi_{\rho}(x)}$; then $\eta(x) \leq 0$ for $x>0$, $\eta(x) \geq 0$
for $x\leq 0$, and $-\eta\pi/2\leq \eta(x)\leq \eta\pi/2 $, $\forall
x\in\R$. Now a simple calculation gives 
\be\label{3.5}
\begin{array}{l}
{\ds \re {V_\rho (x)} = \sin{\left\{\epsilon(2k-1)/(2k+1)\right\}}\left(x^2 -\eta(x)^2
\right) - \cos{\left\{\epsilon(2k-1)/(2k+1)\right\}}\left(2x\eta(x)\right) }\\
{\ds -\rho\eta(x)\left[(2k+1)x^{2k} - \left( \begin{array}{c} 2k+1 \\ 3
\end{array}\right) x^{2k-2}\eta(x)^2 + \left( \begin{array}{c} 2k+1 \\ 5
\end{array}\right)x^{2k-4}\eta(x)^4  \right.}\\
{\ds \left. + ... + (-1)^{k-1}\left( \begin{array}{c} 2k+1
\\ 2k-1 \end{array}\right)x^2\eta(x)^{2k-2} + (-1)^k\eta(x)^{2k}\right]}\; ,
\end{array}
\ee
Next we notice that the term inside the square brackets can be bounded from below by a
constant (independent of $\rho$), and for $x\geq n\geq n_0$, $0<\rho<\rho_0$ we have
$x^2 > \eta(x)^2$, whence
\be\label{3.6}
\re {V_\rho (x)} \geq cn + c' \geq \frac{c_1}{R} + c_2 \, .
\ee
For $x\leq -n$ we still have $x^2 > \eta(x)^2$, and the term inside square brackets in
(\ref{3.5}) can be bounded from above by
$$
Ax^{2k} + B \, ,
$$
for suitable constants $A>0$ and $B\in\R$, independent of $\rho$ and $n$. Thus,
\be\label{3.6bis}
\begin{array}{l}
{\ds \re {V_\rho (x)} \geq \sin{\left[\epsilon(2k-1)/(2k+1)\right]}\left(x^2 -
\eta(x)^2\right)} \\
{\ds - \cos{\left[\epsilon(2k-1)/(2k+1)\right]}\left(2x\eta(x)\right) }
\\ {\ds - \rho\eta(x)(Ax^{2k} + B) }
\end{array}
\ee
Now, if the number $a_k>0$ in Definition \ref{D.3.3} is chosen so that the polynomial
term 
\be\label{3.6ter}
-2x\cos{\left[\epsilon(2k-1)/(2k+1)\right]} - \rho(Ax^{2k} + B)
\ee
attains its (positive) maximum at $\ds x_0 = -\frac{a_k}{\rho^{1/(2k-1)}}$ , estimate
(\ref{3.6}) still holds in the interval $x_0\leq x\leq -n$, if we make the assumption,
not restrictive in this context, that $n\ll\rho^{-2k}$. Finally, notice that at some
point smaller than $x_0$ the term (\ref{3.6ter}) becomes negative and tends to
$-\infty$ as $\rho\to 0^+$, without being compensated by the term
$$
\sin{\left[\epsilon(2k-1)/(2k+1)\right]}\left(x^2 - \eta(x)^2\right)
$$
which behaves as $\ds \frac{\rho^{1/q}}{R}x^2$, if we recall that $\ds \sin{\epsilon} =
\frac{\rho^{1/q}}{R}$. This is the reason why it was necessary to set $\eta(x) = 0$ for
$x\leq x_0 - \eta$. In particular in this region we have
$$
\re {V_{\rho}(x)} = \left(\sin{\left[\epsilon(2k-1)/(2k+1)\right]}\right)x^2 \geq
c\left(\frac{\rho^{1/q}}{R}\right)\left(-\frac{a_k}{\rho^{1/(2k-1)}} - \eta \right)^2
\geq \frac{c_1}{R} + c_2 \; ,
$$
whence the assertion.
\vskip 0.2cm\noindent
>From now on the constant $a_k>0$ in Definition \ref{D.3.3} will be chosen so as to
satisfy Lemma \ref{L.3.5}.
\begin{lemma}
\label{L.3.6}
There exist constants $c_3$,$c_4 >0$ such that 
\be\label{3.7}
\re{\langle u, H_{\rho}u \rangle} \geq c_3 \int_{x_0}^{+\infty} \frac{(1
+x^2)^{\frac{1}{4}}}{x^2 + (1+x^2)^{\frac{1}{2}}} |pu|^2 dx - c_4\|u\|^2 \; ,
\ee
$\forall u\in D(H_{\rho})$, $0<\rho <\rho_0$.
\end{lemma}
{{\it Proof.}}
Set $\omega =
e^{-i(\frac{\pi}{2}-\frac{2k-1}{2k+1}\epsilon)}$. Then we have
\be\label{3.8}
\re{\langle u,H_{\rho}u \rangle} = \re{\int_{-\infty}^{+\infty} \left\{\omega
f_{\rho}^2|pu|^2 + \frac{\omega}{4}(f_{\rho}^2)''|u|^2 + V_{\rho}(x)|u|^2\right\} dx} \;
.
\ee
As for the first term in the right hand side of (\ref{3.8}) we have
\be\label{3.9}
\re{(\omega f_{\rho}^2)} = \sin{\left[\epsilon(2k-1)/(2k+1)\right]}\re{f_{\rho}^2} +
\cos{\left[\epsilon(2k-1)/(2k+1)\right]}\im{f_{\rho}^2}\; .
\ee
For $x\geq x_0$ it is easy to check that 
\be\label{3.10}
\re{f_{\rho}^2} \geq \frac{1}{4} \left( 1 - \eta^2 \frac{(1
+x^2)^{\frac{1}{2}}}{[x^2 + (1+x^2)^{\frac{1}{2}}]^2}\right)
\ee
and
\be\label{3.11}
\im{f_{\rho}^2} \geq \eta \left[\frac{(1+x^2)^{\frac{1}{4}}}{x^2 +
(1+x^2)^{\frac{1}{2}}}\right]
\ee 
whence 
\be\label{3.12}
\re{(\omega f_{\rho}^2)}\geq \eta
\left(\cos{\left[\epsilon(2k-1)/(2k+1)\right]}\frac{(1+x^2)^{\frac{1}{4}}}{x^2 +
(1+x^2)^{\frac{1}{2}}}\right) \; .
\ee
In the region $x\leq x_0 - \eta$ we have $f_{\rho}(x) = 1$, so that 
\be\label{3.13}
\re{(\omega f_{\rho}^2)} = \sin{\left[\epsilon(2k-1)/(2k+1)\right]} \, .
\ee
Now simple calculations allow us to verify that $|(f_{\rho}^2)''|$ is bounded. Moreover
from (\ref{3.5}) it follows that $\re{V_{\rho}(x)}$ is bounded from below in the
interval $(-n_0,n_0)$, and therefore in $\R$ by Lemma \ref{L.3.5}. Now the assertion
follows combining this result with (\ref{3.12}) and (\ref{3.13}).
\begin{corollary}
\label{C.3.7}
\begin{itemize}
\item[(1)]
$
\ds \lim_{\rho \rightarrow 0^+} H_{\rho}u = H_0 u \; , \quad \forall u\in D_1 \, .
$
\item[(2)]
$\Delta' \neq \emptyset$, where
$$
\Delta' = \{z\in\C : z\notin\sigma(H_{\rho})\; {\rm and}\; (z - H_{\rho})^{-1}\;
{\rm is\; uniformly\; bounded\; as}\; \rho\to{0^+}\}\, .
$$
\item[(3)]
$H_{\rho}$ converges strongly to $H_0$ in the generalized sense.
\end{itemize}
\end{corollary}
{{\it Proof.}}
Statement (1) follows from the fact that $\xi_{\rho}(x)\to \xi_0(x)$ as $\rho\to 0^+$,
uniformly on compacts. By Lemma \ref{L.3.6} we have that the numerical range of
$H_{\rho}$ is contained in a right half-plane $\Pi$, and since $H_{\rho}$ has discrete
spectrum, 
$\|(z - H_{\rho})^{-1}\| \leq ({\rm dist}(z,\Pi))^{-1}$, $\forall z\notin\Pi$. Finally
(3) follows from (1) and (2), since $D_1$ is a core for $H_{\rho}$, $\rho\geq 0$ (see
\cite{Kato}, Theorem VIII.1.5).
\begin{corollary}
\label{C.3.8}
Let $\chi\in C_0^{\infty}(\R)$ be the function defined in Theorem \ref{T.2.8}(c), and
again let $\chi_n(x) = \chi(x/n)$, $M_n(x) = 1 - \chi_n(x)$, $\forall n\in\N$. Then
there exists $c_5 > 0$ such that 
\be\label{3.14}
\|[H_{\rho},\chi_n]u\| \leq \frac{c_5}{n^{\frac{1}{4}}}(\|H_{\rho}u\| + \|u\|)
\ee
$\forall u\in D(H_{\rho})$, $ 0\leq\rho <\rho_0$.
\end{corollary}
{{\it Proof.}}
Let $u\in D(H_{\rho})$, $\|u\|=1$, and $\gamma_{2n}$ be the characteristic function of
the interval $[-2n,2n]$. We have
\be\label{3.15} 
[H_{\rho},\chi_n] = \omega [pf_{\rho}^2p,\chi_n] =
\omega\gamma_{2n}\{2in^{-1}f_{\rho}^2\chi'(x/n)p + 2n^{-1}f_{\rho}f_{\rho}'\chi'(x/n) + 
n^{-2}f_{\rho}^2\chi''(x/n)\} \; .
\ee
Now, since $\chi'$, $\chi''$, $f_{\rho}$, $f_{\rho}'$, $f_{\rho}^2$ are all bounded
functions, we have the pointwise estimate
\be\label{3.16}
|[H_{\rho},\chi_n]u(x)| \leq \frac{c}{n}(|u(x)| + |(pu)(x)|) \, .
\ee
Thus, for $\|u\|=1$,
\vskip 0.2cm\noindent 
$\ds \|[H_{\rho},\chi_n]u\| $ 
\vskip 0.1cm\noindent 
$\ds \leq \frac{c'}{n}\left\{\left(\int_{-2n}^{2n} |pu|^2\frac{(1
+x^2)^{\frac{1}{4}}}{x^2 + (1+x^2)^{\frac{1}{2}}}\frac{x^2 + (1+x^2)^{\frac{1}{2}}}{(1
+x^2)^{\frac{1}{4}}} dx\right)^{\frac{1}{2}} + 1\right\}$
\vskip 0.1cm\noindent 
$\ds \leq \frac{c''}{n}\left\{n^{\frac{3}{4}}\left(\int_{x_0}^{+\infty} |pu|^2\frac{(1
+x^2)^{\frac{1}{4}}}{x^2 + (1+x^2)^{\frac{1}{2}}} dx \right)^{\frac{1}{2}} + 1 \right\}$
\vskip 0.1cm\noindent 
$\ds \leq \frac{c_5}{n^{\frac{1}{4}}}\left\{\re{\langle u,H_{\rho}u\rangle} + 1
\right\}$ ,
\vskip 0.2cm\noindent
whence the assertion. Notice that to obtain the second inequality we assumed again,
without loss, $n\ll |x_0|$, while for the last inequality we have used Lemma {3.6}.
\begin{lemma}
\label{L.3.9}
Let the sequences $\rho_m\to 0^+$ and $u_m\in D(H_{\rho_m})$ be given such that
$\|H_{\rho_m}u_m\|$ is bounded, $\|u_m\| = 1$, $u_m\stackrel{w}{\rightarrow} 0$. Then
$\forall n$
$$
\lim_{m\rightarrow\infty} \|\chi_nu_m\| = 0 \; .
$$
\end{lemma}
{{\it Proof.}}
Set $ H_{\rho}' = \omega^{-1}H_{\rho}$ and let $\lambda\in\C-\sigma(H_0')$ be fixed.
Then we have
$$
\|\chi_nu_m\|^2 \leq c\left(\|\chi_nR_0'(H_0' - H_{\rho_m}')u_m\|^2 +
\|\chi_nR_0'(H_{\rho_m}' - \lambda)u_m\|^2\right) \; ,
$$
where $R_0' = (\lambda - H_0')^{-1}$. Now we can proceed as in the proof of Lemma 5 of
\cite{Stark}.
\begin{lemma}
\label{L.3.10}
For any $\lambda\in\C$ there exist $R$, $n_0$, $\delta >0$ such that
$$
d_{n,\rho}(\lambda) \equiv \inf{\{\|(\lambda - H_{\rho})M_nu\| : u\in D(H_{\rho}),
\|M_nu\|=1\}} \geq \delta \; ,
$$
$\forall n>n_0$ ,$\forall \rho\leq \rho_0$.
\end{lemma}
{{\it Proof.}}
By Lemma \ref{L.3.5}
$$
\re {\langle M_nu,V_{\rho_m}M_nu\rangle} \geq \frac{c_1}{R} + c_2 > \delta >0
$$
if $\|M_nu\|=1$ and $R$ is chosen sufficiently small. Finally, from the proof of
Lemma \ref{L.3.6} the kinetic part of $H_{\rho}$ is bounded from below and this proves
the lemma.
\vskip 0.2cm\noindent
{{\it Proof of Theorem \ref{T.3.1}.}} 
 From Corollary \ref{C.3.8} and Lemmas \ref{L.3.9}, \ref{L.3.10} the proof of a theorem
analogous to Theorem \ref{T.2.8} immediately follows, with the operator $K(\rho)$
replaced by
$H_{\rho}$, $\rho\geq 0$. Thus, we can apply Theorem A1 of \cite{Db}, in order to
obtain the following stability result:
\begin{itemize}
\item[(i')]
if $\lambda\notin\sigma(H_0)$ then $(\lambda - H_{\rho})^{-1}$ is uniformly bounded as
$\rho\to 0^+$;
\item[(ii')]
if $\lambda\in\sigma(H_0)$ then $\lambda$ is a stable eigenvalue with respect to the
family $\{H_{\rho}\}_{\rho >0}$.
\end{itemize}
\par\noindent
With an argument analogous to the one used to prove Theorem \ref{T.3.1} we now obtain
the following
\begin{theorem}
\label{T.3.11}  
Let $ q = (2k-1)/{2}$. Then for each eigenvalue $E_j(\beta)$, $j\in\N$, of
$H(\beta)$, $\im{\beta}>0$, there exists $R'>0$ such that $E_j(\beta)$ is analytic in
the Nevanlinna disk of the $\beta^{1/q}$-plane 
$$
D_{R'} = \{\beta\in\C : |\beta^{1/q} - (R/2)e^{i\pi/q}| \leq
R/2\}
$$
contained in the half-plane $-\frac{\pi}{2} + \frac{\pi}{q} < \arg{\beta^{1/q}}
< \frac{\pi}{2} + \frac{\pi}{q}$, with radius $R/2$ and center at
$C=(R/2)e^{i\pi/q}$.
\end{theorem}
\begin{remark}
\label{R.3.12}
{\rm Set $\beta' = \beta e^{-i\pi}$; then, by Theorem \ref{T.3.11}, $E_j(\beta)$
is analytic in the Nevanlinna disk 
$$
C_{R'} = \left\{\beta\in\C : \re{(\beta')^{-1/q}} \geq (R')^{-1}\right\}
$$
of the $(\beta')^{1/q}$-plane.}
\end{remark}
\begin{theorem}
\label{T.3.13}  
For any $j\in\N$, the eigenvalue $E_j(\beta)$ of $H(\beta)$ is Borel summable in the
ordinary sense for $0<\arg{\beta}<\pi$ and in the distributional sense for $\arg{\beta}
= 0$ and $\arg{\beta} = \pi$.
\end{theorem}
{{\it Proof.}}
We will examine only the "singular" cases $\arg{\beta} = 0, \pi$; the others can be
treated in the standard way (see also \cite{CGM} for $\pi/8 < \arg{\beta} <7\pi/8$).
Let us consider first the case $\arg{\beta} = 0$. Then Theorem \ref{T.3.1} allows us to
apply the criterion for the distributional Borel-Leroy sum of order $q$ given in
\cite{DBS}. More precisely, the criterion requires the analyticity of $E_j(\beta)$ in a
disk $C_R = \{\beta : \re{\beta^{-1/q}} \geq R^{-1} \}$, as obtained in Theorem
\ref{T.3.1}, and the well-known estimates for the remainders:
\be\label{3.17}
\left |E_j(\beta) - \sum_{s=0}^{N-1} a_s\beta^s\right | \leq A\sigma^N
\Gamma(qN+1)|\beta|^N\; , \quad \forall N=1,2,...
\ee
uniformly in $C_{R,\epsilon} = \{\beta\in C_R : \arg{\beta^{1/q}}\geq -\pi/2 +
\epsilon\}$, $\forall \epsilon>0$, where the constants $A$ and $\sigma$ may depend on
$\epsilon$, and $\ds \sum_{s=0}^{\infty} a_s\beta^s$ is the \RS\ perturbation expansion
corresponding to $E_j(\beta)$ (see also \cite{Reed-Simon},Vol.IV, for the standard proof
of such estimates). As for the case $\arg{\beta} = \pi$, we first notice that
(\ref{3.17}) is known to hold uniformly in $\beta$ in any sector
$$
S(\delta) = \left\{\beta\in\C : |\beta|<B(\delta),-\frac{\pi}{2}+\frac{\delta}{q}<
\arg{\beta^{1/q}}<\frac{\pi}{2}+\frac{\pi}{q}-\frac{\delta}{q}\right\} \; .
$$
Next observe that the direction $\arg{\beta} = \pi$ in the $\beta$-plane corresponds to
the direction $\arg{\beta'} = 0$ in the $\beta'$-plane, $\beta' =
\beta e^{-i\pi}$. Now, in analogy with \cite{DBS} (Theorems 3 and 4), the
criterion for the distributional Borel-Leroy summability of order $q$ of $E_j(\beta)$
in the direction $\arg{\beta} =
\pi$ can be stated in terms of the "adapted" variable $\beta'$, in the sense that it
relies on the following two conditions:
\begin{itemize}
\item[(1)]
$E_j(\beta)$ is analytic in 
$$
C_{R'} = \left\{\beta\in\C : \re{(\beta')^{-1/q}} \geq (R')^{-1}\right\} \; ;
$$
\item[(2)]
$\forall \epsilon >0$, there exist $A$,$\sigma > 0$ such that
\be
\label{3.18}
\left |F_j(\beta') - \sum_{s=0}^{N-1} (-1)^s a_s(\beta')^{s}\right | \leq A\sigma^N
\Gamma(qN+1)|\beta'|^{N} \; , \quad \forall N=1,2,...
\ee
uniformly in $C_{R',\epsilon} = \{\beta\in C_{R'} : \arg({\beta'})^{1/q}\geq -\pi/2 +
\epsilon\}$, where
$$
F_j(\beta')\equiv \overline{E_j\left(\overline{\beta'} e^{-i\pi} \right)} =
\overline{E_j(\overline{\beta})} \, .
$$
\end{itemize}
Now, (1) is given in Remark \ref{R.3.12} and (2) follows from the fact that the sector
$S(\delta)$, where (\ref{3.17}) holds uniformly,  can be rewritten in terms of
$(\beta')^{1/q}$ as 
$$
S(\delta) = \left\{\beta\in\C : |\beta|<B(\delta),-\frac{\pi}{2}
-\frac{\pi}{q}+\frac{\delta}{q}<
\arg{(\beta')^{1/q}}<\frac{\pi}{2}-\frac{\delta}{q}\right\} \; .
$$
Indeed, since the coefficients $a_s$ of the power series are real, (2) is equivalent to
\be
\label{3.18bis}
\left |\overline{F_j(\overline{\beta'})} - \sum_{s=0}^{N-1} (-1)^s a_s(\beta')^{s}\right
|
\leq A\sigma^N
\Gamma(qN+1)|\beta'|^{N} \; , \quad \forall N=1,2,...
\ee
uniformly in $\overline{C}_{R',\epsilon} = \{\beta\in C_{R'} : \arg({\beta'})^{1/q}\leq
\pi/2 -
\epsilon\}$, where $\overline{F_j(\overline{\beta'})}=E_j(\beta)$.
\vskip 0.2cm\noindent
{\it Proof of Theorem \ref{T.1.1}}
 According to the terminology introduced in \cite{DBS} about the distributional
Borel summability, by (\ref{3.17}) $ E_j(\beta)$ represents the so-called
"upper sum" and
$\overline{E_j(\overline\beta)}$ the "lower sum" for $\beta\in C_R$; conversely, by
(\ref{3.18}), 
$E_j(\beta)$ is the lower sum and $\overline{E_j(\overline\beta)}$ the upper sum for 
$\beta\in C_{R'}$. More precisely, $E_j(\beta)$ admits for $\beta\in C_R$  the integral
representation
\be
\label{uppersum}
E_j(\beta)=\frac{1}{q\beta}\int_0^{\infty}B_j(t+i0)
e^{-(t/\beta)^{1/q}}\left(\frac{t}{\beta}\right)^{-1+1/q}\,dt
\ee 
and the analogous representation holds for $\overline{E_j(\overline\beta)}$ with
$\overline{B(t+i0)}$ in place of $B(t+i0)$. For $\beta \in C_{R'}$ the representation
analogous to (\ref{uppersum}) holds in terms of the adapted variable $\beta'$, i.e.:
\be
\label{uppersumbis}
\overline{E_j(\overline\beta)}={F_j({\beta'})}=
\frac{1}{q\beta'}\int_0^{\infty}B_j(t+i0)
e^{-(t/\beta')^{1/q}}\left(\frac{t}{\beta'}\right)^{-1+1/q}\,dt
\ee 
because the odd terms in the power series are identically zero. The distributional Borel
sum, which must be real for
$\beta\in\R$ since the \RS\ perturbation series $\ds
\sum_{s=0}^{\infty} a_s\beta^s$ has real coefficients, is given by 
\be
\label{3.19}
f_j(\beta) = \frac{E_j(\beta) + \overline{E_j(\overline\beta)}}{2} \; ,
\ee
 while the difference
\be
\label{3.20}
d_j(\beta) \equiv 2ig_j(\beta) = \left\{\begin{array}{l}   E_j(\beta) -
\overline{E_j(\overline\beta)},\quad\beta\in C_R \\
{}
\\
 \overline{E_j(\overline\beta)}-E_j(\beta), \quad \beta\in C_{R'}
\end{array}
\right.
\ee
 represents the so-called "discontinuity", which has zero asymptotic expansion. 
Now, if $\beta\in\R$, by (\ref{uppersum}) and (\ref{uppersumbis}) we have 
$$ 
E_j(-\beta)=\overline{E_j(\beta)}
$$
since, once again, the perturbation series
$\ds
\sum_{s=0}^{\infty} a_s\beta^s$ is such that
$a_s = 0$ if $s$ is odd, and therefore it can be written in the form $\ds
\sum_{l=0}^{\infty} a_{2l}\beta^{2l}$. It follows that $f_j(\beta) = f_j(-\beta)$ and
$g_j(-\beta)=-g_j(\beta)$, i.e. 
$E_j(\beta)$ and $E_j(-\beta)$ have the same real part and opposite imaginary one.
This concludes the proof of the theorem.
\begin{remark}
\label{R.3.14}
\begin{enumerate}
\item{\rm For $\beta\in\R$, it follows from (\ref{3.19}) and (\ref{3.20}) that 
\be
\label{3.21}
f_j(\beta)=\re E_j(\beta), \quad d_j(\pm|\beta|)=\pm 2i\im E_j(\pm|\beta|).
\ee
 Since
$E_j(\beta)$ can be interpreted as a resonance of the problem (\cite{CM}), $f_j(\beta)$
represents the position of the resonance and
$|d_j(\beta)|/2$ its width. As in the Stark effect, the distributional Borel
summability completely determines the resonance.}
\item {\rm 
In
the present case $f_j(\beta)$ and $d_j(\beta)$ admit a further interpretation, since
by Remark
\ref{R.2.10}, $\overline{E_j(\overline\beta)} = E_j^1(\beta)$, where $E_j^1(\beta)$
represents the $j$-th eigenvalue of $H(\beta)$ for $\im{\beta}<0$. As proved for 
$ E_j(\beta)$, $E_j^1(\beta)$ can be analytically continued to Nevanlinna disks
analogous to $C_R$ and $C_{R'}$ across the positive and negative real axis
respectively. Thus,}
$$
f_j(\beta) = \frac{E_j(\beta) + E_j^1(\beta)}{2} \quad {\rm and}\quad d_j(\beta) =
\pm [E_j(\beta) - E_j^1(\beta)] \; ,
$$
{\rm where the $+$ holds for $\beta\in C_R$, and the $-$ for $\beta\in C_{R'}$.}
\item
{\rm As already recalled, the eigenvalues admit the classical Borel integral
representation for $\pi/8 +\eta < \arg{\beta}< 7\pi/8 -\eta, \eta >0$ 
 (\cite{CGM}).  Formulas (\ref{uppersum}), (\ref{uppersumbis}) yield their explicit
analytic continuation to the regions $C_R$ and $C_{R'}$ across the real axis.}
\end{enumerate}
\end{remark}

\end{document}